\documentclass{aa}
\usepackage{natbib}
\usepackage{graphicx}
\begin{document}
\title{The dynamical disconnection of sunspots\\ 
              from their magnetic roots}
\titlerunning{Disconnection of sunspots}
\author{M. Sch\"ussler \inst{1}
\and M. Rempel \inst{2}}
\institute{Max-Planck-Institut f\"ur
Sonnensystemforschung,
Max-Planck-Str. 2, 37191 Katlenburg-Lindau, Germany 
\and High Altitude Observatory,
NCAR, P.O. Box 3000, Boulder, Colorado 80307, USA}


\date{Received; accepted} \abstract{ After a dynamically active
emergence phase, magnetic flux at the solar surface soon ceases to show
strong signs of the subsurface dynamics of its parent magnetic
structure.  This indicates that some kind of disconnection of the
emerged flux from its roots in the deep convection zone should take
place.  We propose a mechanism for the dynamical disconnection of the
surface flux based upon the buoyant upflow of plasma along the field
lines. Such flows arise in the upper part of a rising flux loop during
the final phases of its buoyant ascent towards the surface. The
combination of the pressure buildup by the upflow and the cooling of the
upper layers of an emerged flux tube by radiative losses at the surface
lead to a progressive weakening of the magnetic field in several Mm
depth. When the field strength has become sufficiently low, convective
motions and the fluting instability disrupt the flux tube into thin,
passively advected flux fragments, thus providing a dynamical
disconnection of the emerged part from its roots. We substantiate this
scenario by considering the quasi-static evolution of a sunspot model
under the effects of radiative cooling, convective energy transport, and
pressure buildup by a prescribed inflow at the bottom of the model.  For
inflow speeds in the range shown by simulations of thin flux tubes, we
find that the disconnection takes place in a depth between 2 and 6 Mm
for disconnection times up to 3 days.
}
\maketitle

\section{Introduction}
\label{Sec:intro}

Until the advent of local helioseismology, very little was known about
the subsurface structure of sunspots and active regions.  Differences in
the measured rotation rates of the nonmagnetic plasma and the magnetic
`tracers' were taken as evidence that the latter are somehow `anchored'
in deeper layers, but no consensus could be reached as to where and how
this anchoring should precisely take place
\citep[e.g.,][]{DSilva:Howard:1994,Javaraiah:Gokhale:1997,Beck:2000,
Schuessler:1984,Schuessler:1987}.  The recent results from local
helioseismology indicate significant changes of the thermodynamic and/or
magnetic properties as well as of the flow field at a depth of less than
10 Mm below sunspots \citep{Kosovichev:etal:2000,Zhao:etal:2001,
Couvidat:etal:2004} and active regions
\citep{Basu:etal:2004,Hindman:etal:2004}. These findings are consistent
with the model of a sunspot as a shallow object, which breaks apart into
a large number of flux tubes not far below the surface
\citep{Parker:1979b,Spruit:1981b,Choudhuri:1992}, rather than that of a
monolithic plug of magnetic flux extending deep into the convection zone
\citep[see also the detailed discussion of sunspot models
by][]{Thomas:Weiss:1992}.

From the theoretical side, analytical studies and numerical simulations
of magnetic flux tubes have provided a consistent picture of the
formation of active regions and sunspot groups, comprising the storage
and amplification of magnetic flux near the bottom of the convection
zone as well as the formation of unstable loops and their rise through
most of the convection zone.  Given a strongly super-equipartition
initial field of the order of $10^5\,$G (10~Tesla) at the bottom of the
convection zone, large-scale properties of newly-emerged active regions
and sunspot groups, like low emergence latitudes, systematic tilt angles
(Joy's law), and asymmetric proper motions of the two polarities, can be
quantitatively explained \citep[see reviews by][and further references
therein]{Moreno-Insertis:1997b,Fisher:etal:2000,Fan:2004}.

Numerical studies following the evolution of an unstable flux loop from
its origin at the bottom of the convection zone only reach until about
10 Mm from the surface, where the thin-flux-tube approximation breaks
down. Realistic simulations of the subsequent phases are still too
demanding in view of the computing power available today. Therefore, the
actual emergence of flux in the photosphere and the evolution thereafter
has not been covered by such simulations \citep[except for rather
idealized situations aiming at describing the effects on the coronal
magnetic structure,
e.g.,][]{Fan:Gibson:2004,Magara:2004,Archontis:etal:2004}. On the other
hand, a wealth of observational evidence indicates a remarkable change
in the dynamical properties of sunspots and active regions from `active'
to `passive' evolution shortly after their emergence
\citep{Schuessler:1987, Schrijver:Title:1999}.  The emerging flux
initially displays the clear signature of a rising, fragmented flux tube
and evolves according to its internal large-scale dynamics
\citep[e.g.,][]{McIntosh:1981, Strous:etal:1996}. However, the expansion
of the bipolar region and the proper motion of the sunspots with respect
to the surrounding plasma decays within a few days after emergence and
larger magnetic structures start to fragment into small-scale flux
bundles, which are largely dominated by the local near-surface flows
(granulation, supergranulation, differential rotation, meridional
circulation).  The magnetic flux is then passively transported by these
velocity fields and becomes dispersed over wide areas\footnote{When the
average flux density exceeds a limit of about 100~G, the relatively
stable `plage state' \citep{Schrijver:1987b} is maintained as an
intermediate stage: the convective pattern is severly disturbed by the
magnetic field and supergranulation is largely absent, so that the
dispersal of the flux is temporarily diminished. Sunspots actively shape
the surrounding flow structure by developing the outward moat flow,
which probably temporarily suppresses the flux transport by
supergranulation and meridional flow.}, a process well represented by
the so-called surface transport models
\citep[e.g.,][]{Wang:etal:1989,Ballegooijen:etal:1998a,Schrijver:2001,
Baumann:etal:2004}.  This behavior is in striking contrast to what would
be expected if the emerged flux would follow the evolution of its
magnetic roots at the bottom of the convection zone:

\smallskip
\noindent{\it Longitudinal drift:} The size range of bipolar magnetic
regions corresponds to azimuthal wavenumbers between $m=10$ and 60,
while the magnetic instabilities leading to loop formation favor values
of $m=1$ or 2 \citep{Spruit:Ballegooijen:1982,Schuessler:etal:1994,
Ferriz-Mas:Schuessler:1995}. Consequently, the two polarities of an active
region should move apart in longitudinal direction much further than
actually observed. Moreover, there is no static equilibrium of the
subsurface vertical `legs' of an emerged bipolar region that is
connected to an azimuthal field of $10^5\,$G at the bottom of the
convection zone \citep{Ballegooijen:1982}. The horizontal component of
the magnetic tension force resulting from the bends of the flux tube
where it turns horizontal sustains the drift of the two legs (and their
associated poles at the surface) in opposite azimuthal
directions. Thin-tube simulations in fact clearly show that the two
poles drift apart unrestrained and even move around the whole
circumference of the Sun within a few months. Clearly, no such
systematic motion is observed.

\smallskip
\noindent{\it Poleward drift:} The emergence of a loop cuts off the
azimuthal flow neccessary for mechanical equilibrium of the submerged
field \citep{Moreno-Insertis:etal:1992}. As a result, the magnetic
structure starts to drift poleward in response to the unbalanced
latitudinal component of the magnetic tension force. No such
systematic poleward drift is observed in bipolar regions or sunspot
groups.

\smallskip
\noindent{\it Tilt angle:} The tilt angle of sunspot groups with respect
to the East-West direction arises through the action of the Coriolis
force on a rising and horizontally expanding flux loop
\citep[e.g.,][]{Dsilva:Choudhuri:1993}. After emergence, the expansion
stops and the flux tube should relax back to its original East-West
orientation during the lifetime of a large bipolar region
\citep{Fan:etal:1994}. No such relaxation is observed
\citep{Toth:Gerlei:2004}.

The observed change of the dynamics of magnetic structures from
`active' to `passive' and the points listed above clearly indicate
that bipolar magnetic regions somehow become disconnected from their
deeper roots within a few days after emergence. Moreover, since the
expected motions of sunspots would otherwise be quite marked, the
complete absence of such behaviour indicates that the disconnection
process must reliably work, virtually without exception. This
represents an important constraint.

To our knowledge, two mechanisms for the disconnection of emerged flux
have been proposed so far. \citet{Schrijver:Title:1999} have suggested
subsurface reconnection of the two opposite polarities leading to
disconnected U-loops in the upper layers. It does not seem obvious how
this mechanism can achieve the required perfect reliability in view of
the strong tendency of the poles to move apart in the longitudinal
direction (see Sect.~\ref{Sec:model}). Earlier, \citet{Fan:etal:1994}
had given arguments for the necessity of disconnection and proposed a
mechanism for the `dynamical disconnection' of emerged flux by a local
loss of lateral total pressure balance following the establishment of an
isentropic hydrostatic equilibrium along magnetic field lines.  As
discussed in Sec.~\ref{Sec:model}, the process sketched by
\citet{Fan:etal:1994} does not seem to operate sufficiently rapidly to
provide disconnection within the first few days after emergence.  Here
we propose a variation of the dynamical disconnection scenario, which is
based upon the strong, buoyancy-driven upflow associated with the final
phase of the rise of a flux loop through the convection zone. We discuss
the various scenarios for disconnection in Sect.~\ref{Sec:model} and
provide an illustrative quantitative elaboration of our model in
Sect.~\ref{Sec:explosion}. The results are discussed and put into a
broader perspective in Sect.~\ref{Sec:discussion}.

\section{Scenarios for the  disconnection of emerged flux}
\label{Sec:model}

\citet{Schrijver:Title:1999} have suggested that active regions
literally become disconnected from their roots by a subsurface
reconnection between the opposite polarities in their (fragmented) legs;
this would create shallow O- or U-loops, which then could float freely
with the near-surface velocity fields.  The problem with this concept is
that, as explained above, the two legs of a bipolar region tend to
rapidly move apart from each other owing to the geometry of the rising
loop and as a result of the unbalanced tension force in the very deep
part. Thus is would require a rather organized and converging flow to
bring the opposite fluxes together to reconnect. Furthermore, the more
separated the two legs already are, the more difficult is it to bring
the opposite-polarity flux together. Consequently, once reconnection has
failed to occur in time for whatever reason and the two poles continue
to drift further apart, they cannot be disconnected any more and
separate indefinitely. To our knowledge, not a single example of such a
behaviour of a larger active region has been reported by observers.

\citet{Fan:etal:1994} have suggested that the establishment of
hydrostatic equilibrium along a flux tube could lead to a dynamical
disconnection of the emerged part from its deeper magnetic roots, which
does not explicitely require a change of the magnetic topology (i.e.,
reconnection): as the subsurface part of the emerged flux tube
approaches hydrostatic equilibrium, it eventually loses lateral pressure
equilibrium at a certain height. This leads to a region of weak, passive
magnetic field that effectively decouples the parts of the flux tube
above and below this height.

This mechanism is closely related to the `explosion' of magnetic flux
tubes discussed by \citet{Moreno-Insertis:etal:1995}: a slowly rising
flux loop that maintains approximate hydrostatic equilibrium along the
field lines experiences a sudden catastrophic weakening of the field
strength at its apex when it reaches a critical height.  This {\em
explosion height\/} depends on the field strength in the part of the
flux tube that remains at the bottom of the convection zone. The plasma
is nearly isentropic within the adiabatically rising flux loop whereas
the entropy decreases with height in the surrounding superadiabatically
stratified convection zone.  As a consequence of the higher entropy
within the magnetic flux tube, the internal pressure decreases more
slowly with height than the external pressure and thus eventually both
become equal at a certain height above the bottom of the convection zone
\citep{Schuessler:Rempel:2002}. When the loop apex approaches this
`explosion height', it expands drastically since pressure balance forces
the magnetic field to become very small, so that it is no longer
dynamically relevant \citep{Rempel:Schuessler:2001}.  From mixing-length
models of the solar convection zone one finds that a hydrostatic flux
tube with a field strength of O($10^5$)~G at the base of the convection
zone should explode at a depth of less than 10~Mm below the surface.
On the other hand, dynamical simulations show that rising
flux loops with such initial field strengths actually traverse this
height range without exploding because a flow along the field lines
keeps them sufficiently far away from hydrostatic equilibrium
\citep[e.g.,][]{Caligari:etal:1995}.

The problem with the scenario of \citet{Fan:etal:1994} is the timescale
for the establishment of the `global' hydrostatic equilibrium required
for the disconnection. The associated time scale if of the order of the
Alfv{\'e}n travel time along the circumference of the flux tube: taking
an Alfv{\'e}n speed of $1\,$km$\cdot$s$^{-1}$ (corresponding to
$10^5\,$G at the bottom of the convection zone), this corresponds to
about a month, much too long to avoid an unrealistic evolution of the
emerged part.

However, a more detailed consideration of the development in the upper
parts of an emerged flux tube reveals a much faster and more local route
towards dynamical disconnection, avoiding the requirement of a global
hydrostatic equilibrium.  Consider the following situation. A rising
flux loop has emerged at the surface to form a bipolar magnetic
region. The subsurface part of the loop has not exploded at its formal
explosion height because the strongly buoyant and rapidly rising plasma
in the loop maintains a super-hydrostatic gradient of the gas pressure
(i.e., a steeper upward decline of the pressure than in the hydrostatic
case) connected with an accelerated upflow \citep{Caligari:etal:1995}.
After emergence, the near-surface parts of the sunspots, pores, and
other flux concentrations are rapidly cooled by radiative losses, which
gives rise to an inward propagating cooling front accompanied by a local
downflow. This leads to a decrease of the gas pressure and a concomitant
intensification of the magnetic field in the upper layers of the flux
tube, while the upflow from below and the downflow from above increase
the gas pressure below the first few Mm depth.  After a few days, the
gas pressure has increased sufficiently to approach the ambient pressure
somewhere between 2 and 10~Mm depth. A thin flux tube (requiring lateral
balance of total pressure) would then lose its equilibrium and explode.
A more general magnetostatic equilibrium could possibly be maintained
for some time but, in any case, the magnetic field becomes strongly
weakened. Once the field strength falls below the equipartition value
with respect to the external convective motions, the field becomes
largely passive in that height range and progressively fragments, so
that the upper, magnetostatic part becomes dynamically disconnected from
its roots.

\section{A simple model for dynamical disconnection}
\label{Sec:explosion}

In order to evaluate the explanatory potential of the scenario sketched
above and to provide an illustrative example, we study the quasi-static
evolution of a sunspot model very similar to that of
\citet{Deinzer:1965}.

\subsection{Model description}
 
\subsubsection*{Evolution}
To realistically follow the sequence of events from flux tube emergence
to disconnection would require a fully dynamic realistic numerical
simulation, which is beyond the scope of this paper. We restrict
ourselves to sequences of quasi-static models of a sunspot and its
underlying flux tube, which evolve in reaction to surface cooling and to
the increasing pressure resulting from an inflow through the
lower boundary of the model. Such an approach is justified as long as
the upflow velocity is significantly smaller than the Alfv{\'e}n speed,
so that the magnetic structure remains almost unaffected by the
dynamical pressure of the flow.  Starting from an isentropic initial
state, we proceed in constant intervals of time and, for each time step,
determine the entropy change due to radiative and convective energy
transport. Next we integrate the gas pressure, taking into account the
growing base pressure owing to the assumed upflow, apply the
corresponding adiabatic correction to the temperature, and calculate the
modified magnetostatic equilibrium. Then we move to the next
timestep. In the following subsections we explain the various
ingredients of the model in more detail.

\subsubsection*{Magnetostatic equilibrium}

We basically follow the approach of \citet{Deinzer:1965} and consider a
similarity solution \citep{Schlueter:Temesvary:1958} for an
axisymmetric magnetostatic configuration with hydrostatic equilibrium
along the field lines and a mixing-length model of the convection zone
\citep{Kiefer:etal:2000} representing the (fixed) external
stratification.

The similarity model assumes an axisymmetric magnetic structure
with a self-similar profile of the magnetic field as a function of the
radial coordinate, $r$, in cylindric coordinates. The two components
of the (untwisted) magnetic field are written in the form
\begin{eqnarray}
  B_z(r,z) &=& f(\zeta)\, B_0(z)\,,
\label{bz} \\
\noalign{\vskip 2mm}
  B_r(r,z) &=& -{r\over 2}\,f(\zeta)\,{\partial B_0(z)\over\partial z}\,,
\label{br}
\end{eqnarray}
where $B_0(z)$ is the magnetic field along the symmetry axis and
$\zeta = r\sqrt{B_0(z)}$. The function $f(\zeta)$, which describes the
radial profile of the vertical field component, can be freely
chosen. We follow the usual practice and use a Gaussian:
$f(\zeta)=\exp(-\zeta^2)$. Inserting the ansatz given by Eqs.~(\ref{bz})
and (\ref{br}) into the equations for magnetostatic equilibrium,
\begin{eqnarray}
0 &=& -\frac{\partial p}{\partial r} + {B_z\over 4\pi}\left(
    {\partial B_r\over\partial z} - {\partial B_z\over\partial r}
    \right)\,,  \\
\noalign{\vskip 2mm}
0 &=& -{\partial p\over\partial z} - {B_r\over 4\pi}\left(
    {\partial B_r\over\partial z} - {\partial B_z\over\partial r}
    \right) - \rho g\;, 
\label{magnetostat}
\end{eqnarray}
and integrating the first of these over $r$ (from 0 to $\infty$) for
constant $z$, yields the ordinary differential equation
\begin{eqnarray}
   \frac{\Phi}{2\pi}\,y\frac{{\rm d}^2 y}{{\rm d}z^2}&=&y^4
   -8\pi(p_{\rm e}-p_{{\rm i}})
   \label{bfield},
\end{eqnarray}
where $y=\sqrt{B_0(z)}$, $\Phi$ is the total magnetic flux, $p_{\rm
e}(z)=p(\infty,z)$ is the (fixed) external pressure, and $p_{{\rm
i}}=p(0,z)$ is the pressure on the axis. The latter is simply given
by hydrostatic equilibrium along the axis, viz.
\begin{equation}
   \frac{{\rm d} p_{{\rm i}}}{{\rm d}z} 
    = \varrho_{{\rm i}}\,g \,,
\label{hydro}\\
\end{equation}
where $\varrho_{{\rm i}}(z)$ is the density profile along the axis and
$g(z)$ is the (depth-dependent) gravitational acceleration.
Consequently, the self-similarity assumption reduces the determination
of the magnetohydrostatic force balance to solving two ordinary
differential equations, namely hydrostatic equilibrium along the central
field line of the axisymmetric configuration and an equation relating
the field strength on the axis to the difference between the gas
pressure on the axis and the (external) pressure far away from the axis.
Note that the thin flux tube approximation with $B^2=y^4=8\pi(p_{\rm
e}-p_{{\rm i}})$ is recovered in the limit $\Phi\rightarrow 0$. The
left-hand side of Eq. (\ref{bfield}) represents the contribution of
magnetic stress integrated from the axis of the field configuration to
infinity, which is not considered in the thin flux tube limit. A
detailed derivation of Eq. (\ref{bfield}) can be found in
\citet{Schlueter:Temesvary:1958}.

We solve Eq. (\ref{bfield}) through an iterative relaxation
procedure,
\begin{equation}
   y_{n+1}=y_n+\frac{\varepsilon}{y_n^3}\left[\frac{\Phi}{2\pi}
     y_n \frac{{\rm d}^2 y_{n+1}}{{\rm d}z^2}
     -y_n^3 y_{n+1}+8\pi(p_{\rm e}-p_{{\rm i}})\right],
\end{equation}
where $y_n(z)$ is the $n$th iteration of the solution. This procedure
avoids numerical stiffness problems introduced by the term $y^4$ and
allows us to specify boundary conditions for $y$ on both sides of the
computational domain.  The prefactor $y_n^{-3}$ has been introduced in
order to  increase the convergence and stability of the algorithm, which
we optimize by the choice of the free parameter $\varepsilon$.

\subsubsection*{Energy transport}
We consider vertical radiative transfer in the grey diffusion
approximation. Since the energy flux of a sunspot umbra cannot be
provided by radiation alone, a reduced level of convective energy
transport is required in addition \citep{Deinzer:1965}. The phenomenon
of umbral dots is often taken as a manifestation of small-scale
convection in the strong umbral magnetic field. In the absence of a
better theoretical model, we treat the magneto-convective energy
transport using a mixing-length approach, taking into account the
inhibiting effect of the magnetic field through a reduction of the
mixing-length parameter. Horizontal radiative energy exchange can be
neglected in the case of sunspots, which are much larger than the photon
mean free path of the order of 100~km at optical depth unity.  The time
evolution of $T_{\rm i}$ then follows from
\begin{equation}
  \varrho_{{\rm i}} c_p \frac{\partial T_{\rm i}}{\partial t}=-\frac{\partial}{\partial z}
  (F_{\rm rad}+F_{\rm conv}),\label{thermo}
\end{equation}
where the radiative energy flux is given by
\begin{equation}
  F_{\rm rad}=-\frac{16\sigma T_{\rm i}^3}{3 \kappa_R \varrho_{{\rm i}}}\,
  \frac{\partial T_{\rm i}}{\partial z}.
\end{equation}
The Rosseland mean opacity, $\kappa_R(\varrho_{{\rm i}},T_{\rm i})$, is
interpolated from a table. Since the value of $\kappa_R$ significantly
drops in the surface layers, the radiative heat conductivity reaches
large values, so that we use a semi-implicit treatment of the radiative
energy flux in order to avoid a too severe time step constraint.

The convective energy flux along the flux tube is determined following 
 \citet{Spruit:1974},
\begin{equation}
  F_{\rm conv}=-b\sqrt{a} \left(\frac{\mathcal{R}}{\mu}\right)^{1/2}
  \left(\frac{l}{H_p}\right)^{1/2}\varrho_{{\rm i}} c_p T_{\rm i}^{3/2}
  \left(\nabla-\nabla^{\prime}\right)^{3/2}
\end{equation}
where $\nabla$ is the logarithmic temperature gradient with respect to
the logarithm of pressure, $\mu$ is the average molecular weight,
${\mathcal{R}}$ the gas constant, and $l/H_p$ the (constant) mixing
length parameter as fraction of the local pressure scale height.  The
logarithmic temperature gradient gradient $\nabla^{\prime}$ reflects
radiative energy exchange of the convective parcels and follows from
\begin{equation}
  \nabla^{\prime}=\nabla_{\rm ad}-2u^2+2u\left(\nabla-\nabla_{\rm ad}+u^2
  \right)^{1/2}\;,
\end{equation}
with
\begin{equation}
  u=\frac{1}{f\sqrt{a}}\left(\frac{l}{H_p}\right)^{-2}
  \frac{12\sigma T_{\rm i}^3}{c_p\varrho_{{\rm i}}\kappa_R H_p^2}\left(\frac{H_p}{g}
  \right)^{1/2}\;.
\end{equation}
Here $\nabla_{\rm ad}$ denotes the adiabatic logarithmic temperature
gradient and $\sigma$ is Stefan's radiation constant.  The values of the
geometric parameters used are $a=1/8$, $b=1/2$, and $f=3/2$.

In the case of a very strong systematic vertical flow with a time
scale comparable to the turnover time of the convective motions, the
mixing-length approach may be invalidated and, in addition, the
advective entropy transport by the large-scale flow may become
relevant. In our simulations, such a situation occurs only during the
initial cooling phase when a rather strong transient downflow develops
in the first 1--2 Mm depth (see the subsection on the implied vertical
velocity below). For the later evolution of the configuration, the
mixing length model appears to be adequate for the purposes of an
illustrative model.

\subsubsection*{Equation of state}
We include the partial ionization of H, He, and He$^+$. The
values of density, heat capacity, and adiabatic temperature gradient as
functions of pressure and temperature are determined in the course of
the calculation by interpolation in a pre-compiled table.

\subsubsection*{Boundary conditions}
For the temperature, we have to specify boundary conditions at both
ends of the integration domain. At the upper boundary ($z=0$, taken to
be the level of Rosseland optical depth unity of the external
stratification) we fix the temperature to a value (typically $3500$ K)
that is lower than the temperature expected at the level $\tau=2/3$ in
the sunspot. At the bottom, we keep the temperature gradient
adiabatic.

For solving Eqs.~(\ref{hydro}) and (\ref{bfield}) we specify a boundary
condition for $p_{{\rm i}}$ at the bottom of the integration domain and
values for $y$ at the bottom and at the top, respectively.  The magnetic
field strength is fixed at the upper boundary (typically at a value of
$2000\,$G), while the thin flux tube relation, $B=\sqrt{8\pi(p_{\rm
e}-p_{{\rm i}})}$, is assumed to hold at the bottom.

We integrate the hydrostatic balance upward from the lower boundary,
specifying the value of the base pressure for each time step according
to the influx of mass due to the assumed upflow. The effect of lateral
expansion or contraction of the configuration is taken into account in
the calculation of the base pressure. We can alternatively specify a
constant upflow velocity, $v_0$, or a constant total mass flux through
the lower boundary of our integration domain.

In the thin flux tube approximation, the total mass within a flux tube 
containing the (constant) magnetic flux $\Phi$ is given by
\begin{equation}
  m=\Phi\int_0^{z_0}\frac{\varrho_{{\rm i}}}{B} {\rm d} z,
  \label{mtot}
\end{equation}
where ${z_0}$ represents the lower boundary of the integration domain.
The base pressure, $p_{{\rm i}}(z_0)$, is adjusted each time step such
that the change in total mass, $\Delta m$, reflects the inflow of material
across the boundary,
\begin{equation}
  \Delta m=A(z_0)\varrho_{{\rm i}}(z_0) v_0\Delta t=
  \Phi\frac{\varrho_{{\rm i}}(z_0)}{B(z_0)}v_0\Delta t\;,
\end{equation}
where $A(z)=\Phi/B(z)$ is the cross-sectional area of the flux tube.
These relations also hold more generally for the self-similar solutions
considered here, since the profile of the magnetic field in radius is
the same at each depth, $z$, except for a scaling factor.  Therefore, we
can define an `effective' cross section of the configuration at each
height on the basis of the axial field as $A_{\rm eff}(z)=\Phi/B(z)$,
where $A_{\rm eff}$ depends on the radial field profile
considered. $A_{\rm eff}$ is in general different from the geometric
diameter of the flux configuration (for instance, in the case of a
Gaussian profile it formally has infinite extent). However, since the
total mass of the flux tube and the mass flux across the lower boundary
scale in the same way with the cross-section, the difference between
$A_{\rm eff}$ and the geometric cross-section has no influence on the
solution. 

A change of the pressure at the base, $p_{{\rm i}}(z_0)$, alters the mass
within the flux tube through a change of $\varrho_{{\rm i}}$ and $B$ (the
latter is the dominant contribution since $\beta=8\pi p_i/B^2\gg 1$
except for the uppermost layers), which can be written formally as
$m=m[p_{{\rm i}}(z_0)]$. The adjustment of the base pressure $\Delta
p_{{\rm i}}(z_0)$ has to be determined such that
\begin{equation}
  m[p_{{\rm i}}(z_0)+\Delta p_{{\rm i}}(z_0)]=m[p_{{\rm i}}(z_0)]+
  \Phi\frac{\varrho_{{\rm i}}(z_0)}{B(z_0)}v_0\Delta t
  \label{mass_balance}
\end{equation}  
holds. Since the relation between $\Delta p_{{\rm i}}(z_0)$ and $\Delta m$
is non-linear [mainly through Eq.~(\ref{bfield})], we use a Newton
iteration to determine the adjustment of the base pressure such that
Eq.~(\ref{mass_balance}) is satisfied. Note that each iteration step
requires a full integration of the 
Eqs.~(\ref{hydro}) and (\ref{bfield}).

For reasons of consistency with the self-similar solution it is required
that the flow velocity has no variation over the cross section of the
flux tube. This ensures that the magnetic field profiles remain self
similar at all times.

\subsubsection*{Implied vertical velocity}
The inflow at the bottom, the radiative and convective cooling, and the
resulting changes of the magnetostatic balance lead to a net change as
well as to a redistribution of mass within the magnetic structure. 
In our quasi-static model, we can determine the implied vertical
velocity on the axis by considering the equation of continuity:
\begin{equation}
  \frac{\partial}{\partial t}\frac{\varrho_{{\rm i}}}{B}+\frac{\partial}{\partial z}
  \left(v\frac{\varrho_{{\rm i}}}{B}\right)=0\;.
\end{equation}
Starting with $v=0$ at the upper boundary, integration along the tube axis
leads to:
\begin{equation}
  v(z)=-\frac{B(z)}{\varrho_{{\rm i}}(z)}\frac{\partial}{\partial t}\int_0^z
    \frac{\varrho_{{\rm i}}}{B}{\rm d}z^{\prime}\;.
\end{equation}
We determine the time derivative as a finite difference from two
consecutive time steps of the hydrostatic solutions. The feedback of the
dynamic pressure on the solution can be neglected as long as
$\varrho_{{\rm i}} v^2/2 \ll p_{\rm e}-p_{{\rm i}}$.
Even for the maximum inflow velocities that we consider (of the order of
$1000$ m/s), the dynamic pressure is more than two orders of magnitude
smaller than the difference between the external and internal gas
pressures. In other words, the inflow velocity is at least one order of
magnitude smaller than the initial Alfv\'en speed. In all cases that we
have considered, the hydrostatic approximation continues to be valid
throughout the whole simulation.

We have not considered the advection effects from the implied
vertical velocity in the energy equation. In the deeper, homentropic part
of the flux tube, such effects are irrelevant for the (adiabatic)
temperature structure, while the downflow in the upper part could be
temporarily important, but rapidly diminishes to less than
$20\,$m$\cdot$s$^{-1}$ after the first phase of radiative cooling. The
advection of low entropy by the downflow probably enhances the
intensification of the magnetic field in the upper part (in effect, this
is the convective collapse mechanism). A quantitative evaluation of
these effects requires a fully dynamic simulation, which is beyond the
scope of this paper.

\subsubsection*{Initial condition}
We initialize the problem with an isentropic stratification, using an
entropy value equal to the entropy of external stratification at the
lower boundary.  Since the external stratification is strongly
superadiabatic in the uppermost layers of the convection zone, the
magnetic field strength at the lower boundary has to be chosen fairly
large (around $100$ kG for a bottom depth of 12.5 Mm) in order to to
avoid the loss of lateral pressure balance (i.e., an explosion) already
in the initial state. This is a consequence of the magnetostatic
approach chosen here. Therefore, the time evolution in the deeper layers
as calculated here does not fully represent the real solar situation. On
the other hand, the final state of an (nearly) exploded configuration is
a fully consistent and relevant magneto-hydrostatic solution, which
shows that a static sunspot is possible after dynamical disconnection.

\subsection{Numerical results}

\subsubsection*{Example run}

\begin{figure}
  \centering 
  \resizebox{\hsize}{!}{\includegraphics{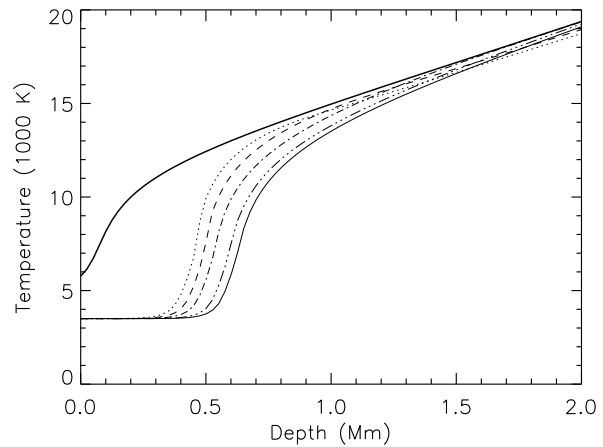}}
\caption{Sequence of temperature profiles as a function of depth in the
upper part of the model, showing the rapid inward propagation of a
cooling front and the establishment of a largely stationary profile
after a few hours. Zero depth corresponds to optical depth unity in the
external stratification. The uppermost (thick full) curve shows the
(time-independent) external temperature. The other curves show the
temperature on the sunspot axis after $0.5\,$h (dotted curve), $1\,$h
(dashed curve), $4\,$h (dash-dotted curve), $15\,$h (dash-triple-dotted
curve), and $30\,$h (full curve), respectively, from the onset of
radiative cooling (corresponding to the emergence of the flux tube at
the surface).  }
\label{fig:evolution_t}
\end{figure}

\begin{figure}
  \centering 
  \resizebox{\hsize}{!}{\includegraphics{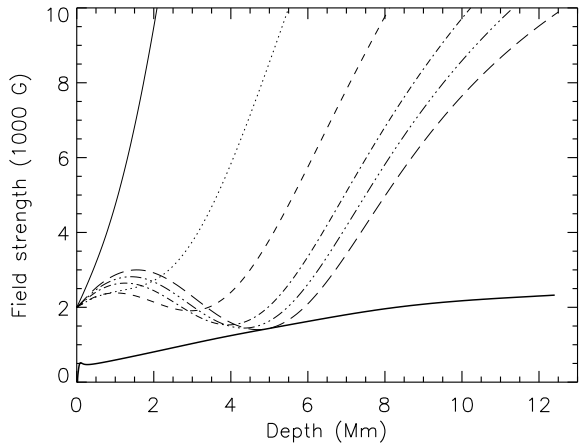}}
\caption{Snapshots from the time evolution of the magnetic field
strength along the tube axis as a function of depth for the same case as
shown in Fig.~\ref{fig:evolution_t} (note the different depth
range). The curves correspond to about $1\,$h (upper full curve),
$10\,$h (dotted curve), $20\,$h (short-dashed curve), $30\,$h
(dash-dotted curve), $35\,$h (dash-triple-dotted curve), and $40\,$h
(long-dashed curve) after the start of radiative cooling
(emergence). The lower, thick full curve gives the equipartition field
strength with respect to the convective velocities from the mixing
length model of the external stratification \citep{Kiefer:etal:2000}.
After an initial general drop, the field strength increases again down
to a depth of about 4 Mm while a local minimum of the field strength
develops and moves downward until, after about 40 hours, the field
strength falls below the local equipartition value at at depth of 4.7 Mm.}
\label{fig:evolution_b}
\end{figure}

\begin{figure}
  \centering 
  \resizebox{\hsize}{!}{\includegraphics{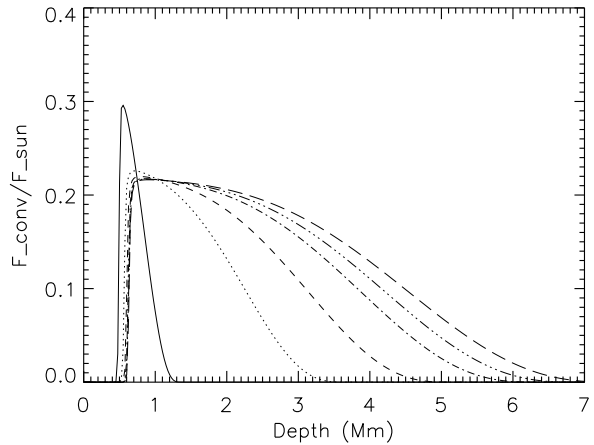}}
\caption{Profiles of the convective energy flux density (normalized by
the undisturbed solar value) along the sunspot axis for the same
instants of time as shown in Fig.~\ref{fig:evolution_b} (note the
different depth range). Above the Wilson depression, the energy
transport is taken over by the radiative flux (not shown here). The
convective transport affects progressively deeper layers in order to
supply the (almost constant) surface energy flux density of about 22\%
of the undisturbed solar value. The corresponding downward extension of
the superadiabatic stratification leads to a growth of the region that
is stabilized against disconnection through the resulting increase of the
magnetic field (cf. Fig.~\ref{fig:evolution_b}. }
\label{fig:evolution_fci}
\end{figure}

As an illustration of a typical evolution in the framework of our
quasi-static model, Figs.~\ref{fig:evolution_t}--\ref{fig:evolution_fci}
show snapshots of temperature, magnetic field strength, and convective
energy flux, respectively, along the axis of the flux tube.  In this
case we have assumed a total magnetic flux of $10^{21}\,$Mx and a
constant mass flux at the bottom (located at $12.5\,$Mm depth)
corresponding to an initial inflow velocity of
$700\,$m$\cdot$s$^{-1}$. Simulations of thin flux tubes give rising
speeds of that order at the same depth \citep{Caligari:etal:1995}. The
spacing of the numerical grid is 25~km.  The temperature at the upper
boundary (at optical depth unity of the external stratification) has
been fixed at a value of 3500~K and the magnetic field strength at
2000~G.  The ratio $\alpha= l/H_p$ of the mixing length, $l$, to the
local pressure scale height, $H_p$, determines the total energy flux in
the flux tube, which is rather well constrained by sunspot observations
at 20--25\% of the undisturbed solar flux.  A value of $\alpha=0.3$
leads to an energy flux of about 22\% of the undisturbed solar value at
the top of the model. About 40 hours after the start of radiative
cooling and bottom inflow, the field strength has fallen below the
equipartition value (with respect to the convective velocity in the
exterior) at at depth of 4.7~Mm, implying the dynamical disconnection of
the upper part from its roots. At the same time, the inflow velocity has
decreased to less than $50\,$m$\cdot$s$^{-1}$ owing to the strong
weakening of the field and the assumed constant mass flux.

Fig.~\ref{fig:evolution_t} shows the fixed external temperature
stratification (thick full line) and five profiles of the temperature in
the upper part of the model at various instants of time. After the
emergence of the flux tube at the photosphere (start of the
calculation), radiative cooling leads to the development of a cooling
front, which rapidly progresses inward. Already after about 30 minutes,
the level of optical depth unity in the flux tube (the Wilson
depression) has reached a depth of about 375 km and after a few hours
the inward propagation of the profile has become very slow (dash-dotted
curve, after 4~hours). The last curve (thin full line) shows the largely
stationary temperature profile about 30~hours after the start of
radiative cooling.  The Wilson depression has reached a value of about
550~km with a temperature of about 4100~K at (Rosseland) optical depth
unity.

For the same case, Fig.~\ref{fig:evolution_b} shows snapshots of the
magnetic field profiles along the tube axis. The full curve shows the
situation after about 1~hour with a steep inward rise of the field
strength that reflects the not quite realistic initial condition. As
time progresses, the depth gradient of the magnetic field strength
flattens and a local minimum develops, which becomes deeper and moves
downward while, at the same time, there is a slow increase of the field
strength in the upper layers. Both effects are connected to the
development of a superadiabatic stratification in the sunspot model,
which results from the growth of the region with convective energy
transport. This is illustrated in Fig.~\ref{fig:evolution_fci}, which
shows depth profiles of the convective energy flux density (normalized
by the undisturbed solar value) for the same instants as the magnetic
field profiles in Fig.~\ref{fig:evolution_b} (note the different depth
scale). The superadiabatic stratification in the region with significant
convective energy flux leads to a decrease of the pressure scale height
and an associated increase of the field strength and thus prevents the
field from becoming weak in the upper layers. As the superadiabatic
region grows downward, the position of the field strength minimum also
is shifted downward until the field strength eventually falls below the
local equipartition level at a depth of about 4.7 Mm. The field strength
at the bottom of the model has then reached about 10~kG, which is consistent
with the results of thin flux tube simulations. The observable layers
above the Wilson depression (optical depth unity, where the field
strength remains around 2500~G) are largely unaffected by this whole
development.

\subsubsection*{Dependence on parameters}

\begin{figure}
  \centering \resizebox{0.8\hsize}{!}{\includegraphics{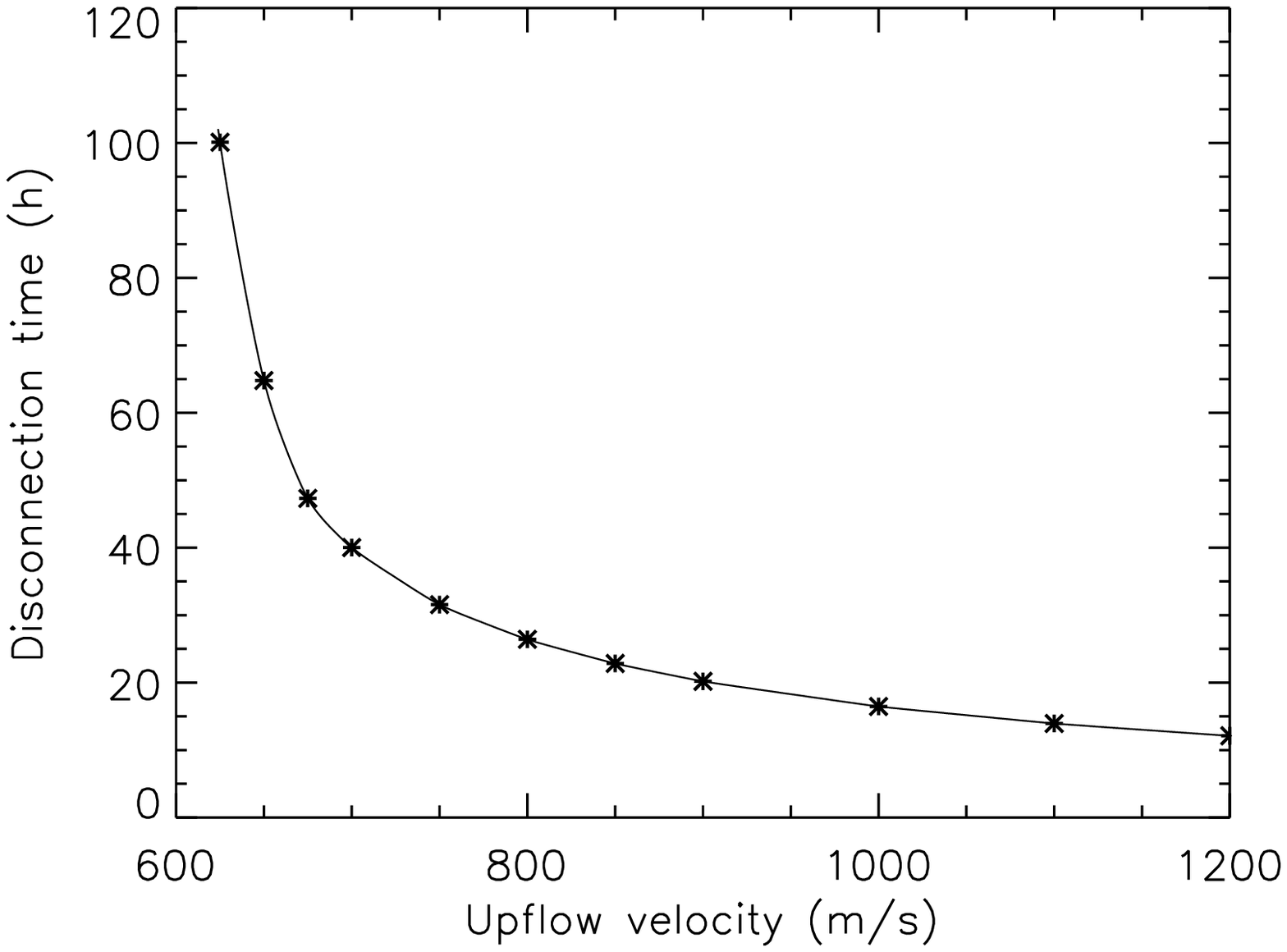}}
  \resizebox{0.8\hsize}{!}{\includegraphics{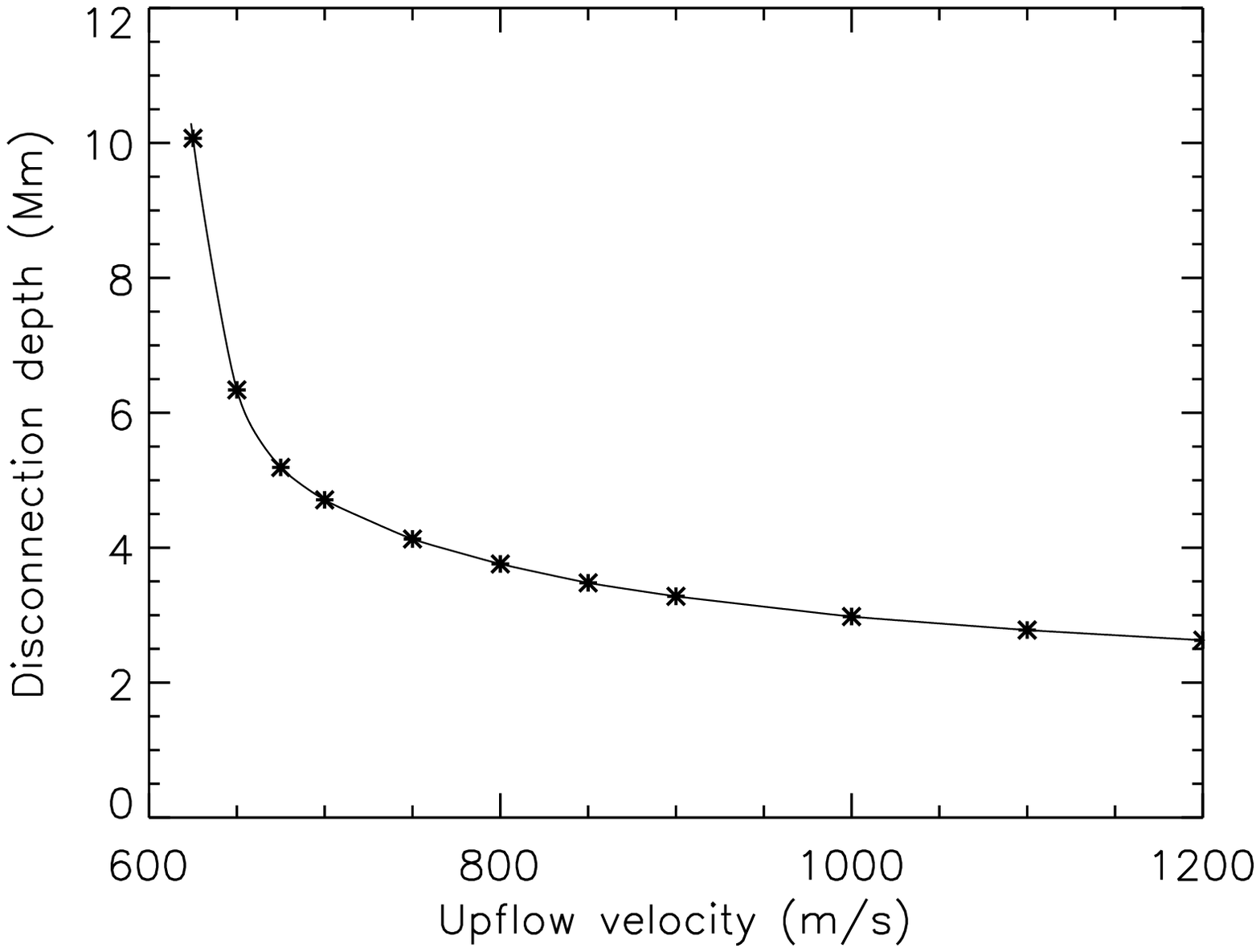}}
  \caption{Disconnection time (upper panel) and disconnection depth
  (lower panel) as functions of the initial upflow velocity at the lower
  boundary. The mass influx is kept constant in time in the individual
  simulation runs (indicated by asterisks). The connecting curves
  represent spline interpolations.}
\label{fig:t_z_exp}
\end{figure}

\begin{figure}
  \centering \resizebox{\hsize}{!}{\includegraphics{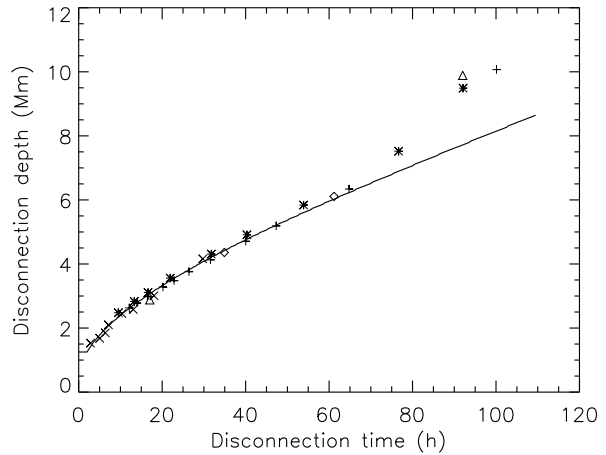}}
  \caption{Disconnection depth vs. disconnection time for various model
  runs. Plus signs: runs with constant mass influx (the same runs as in
  Fig.~\ref{fig:t_z_exp}); asterisks: runs with constant inflow velocity
  in the range 75 -- 400 m$\cdot$s$^{-1}$; diamonds: runs with different
  values for the initial pressure at the top, changing the initial mass
  in the flux tube by about $\pm10$\%; triangles: runs with different
  values of the total magnetic flux ($10^{20}\,$Mx and
  $3\cdot10^{21}\,$Mx, respectively); crosses: runs for which the
  magnetic field was determined according to the pressure difference
  (approximation of thin flux tubes). The curve illustrates the downward
  progression of the zone with convective energy transport
  (cf. Fig.~\ref{fig:evolution_fci}); it shows, as a function of time,
  the depth at which the convective energy flux density equals 10\% of
  the undisturbed solar value, about half of the surface flux density of
  the model sunspot. Except for the largest disconnection depths (which
  may be affected by the proximity of the lower boundary of the model at
  12.5~Mm), the points closely follow the curve, indicating that the
  disconnection depth is mainly determined by the downward progression
  of the convective cooling region.}
\label{fig:points}
\end{figure}

The simulation run presented in the previous subsection yields values
for the time scale and the depth of the disconnection that appear to be
reasonable in view of what is indicated by observations of solar
magnetic regions.  How do these results depend on the various parameters
and assumptions entering the model?

The most important parameter of the model is the inflow velocity at the
bottom, which determines the rate of pressure buildup at the base. For
the case of constant mass flux through the bottom, Fig.~\ref{fig:t_z_exp}
gives the disconnection time and depth (determined as
time and depth for which the field strength first equals the external
equipartition value) as functions of the initial inflow velocity, all
other parameters being the same as those of the reference run discussed
above. The plots show that that smaller values of the mass flux
(smaller initial inflow velocity) lead to later and deeper
disconnection. In the alternative case with constant inflow velocity (as
opposed to constant mass influx), even much smaller inflow speeds of the
order of $100\,$m$\cdot$s$^{-1}$ lead to disconnection within few days
and in a few Mm depth. 

For high initial inflow speeds, the disconnection depth and time become
largely independent of speed and reach values around 9 hours and
2.3~Mm, respectively.  This saturation results from the low
entropy of the gas in the upper layers of the model due to the strong
surface cooling, so that a strong field is always maintained in these
layers. In fact, switching off the radiative losses invariably leads to
a dramatic weakening of the magnetic field at the very top of the
integration domain (depth zero).

We have varied a number of other parameters as well as initial and
boundary conditions in order to evaluate their effects on the
disconnection depth ($z_{\rm d}$) and disconnection time ($t_{\rm d}$).
We find that the results are rather insensitive to the choice of
temperature and magnetic field at the top ($z=0$). The same is true for
the (constant) value of the entropy in the initial stratification. In
most cases we have taken the initial entropy equal to that of the
external medium at the bottom of the model (12.5~Mm depth). Since the
plasma in the flux tube originates from deeper layers zone, its entropy
could be somewhat larger. However, since nearly all of the entropy drop
through the convection zone takes place above 12.5~Mm, even taking the
entropy value corresponding to the bottom of the convection zone has
almost no effect on the results.

However, other parameters and conditions have a significant effect on
the disconnection.  It turns out that $z_{\rm d}$ and $t_{\rm d}$
decrease\\
--  with decreasing depth of the lower boundary,\\
--  with decreasing total magnetic flux,\\
--  with increasing initial mass content,\\
-- and also by keeping the inflow velocity (in contrast to the mass influx)
constant in time.\\
Likewise, determining the magnetic field by replacing the similarity
solution by the thin flux approximation [taking the limit $\Phi\to 0$ in
Eq.~(\ref{bfield})] leads to more shallow and earlier disconnection.

The quantitative results are combined in Fig.~\ref{fig:points}, which
gives $z_{\rm d}$ as a function of $t_{\rm d}$ for a number of cases.
Although the various simulation runs represent quite different
conditions and assumptions, the results suggest a universal relationship
between disconnection time and disconnection depth. The curve drawn in
the figure is not a fit but gives, as a function of time, the depth at
which the convective energy flux density equals 10\% of the undisturbed
solar value, about half of the emergent surface flux density. This curve
therefore reflects the downward progression of the region of convective
cooling as a result of the radiative losses at the surface (see
Fig.~\ref{fig:evolution_fci}).

The relationship between the disconnection depth and the downward
progressing convective `cooling front' is in accordance with our
interpretation of the disconnection depth in the case discussed in the
previous subsection. Without surface cooling, the strongest weakening
of the field would occur near to the top of the model. The cooling
leads to the development of a superadiabatically stratified layer
connected with a strengthening of the magnetic field in the upper part
of the tube, which grows downward as the surface flux has to be
supplied by the thermal energy stored in the deeper layers. This
prevents the field in the upper layers from being weakened by the
pressure buildup due to the inflow from below. The disconnection then
takes place somewhere in the flank of the downward progressing
convective flux profiles shown in Fig.~\ref{fig:evolution_fci},
leading to the relation between $z_{\rm d}$ and $t_{\rm d}$ apparent
from Fig.~\ref{fig:points}. In the absence of convective energy
transport, the cool region would only extend down to the Wilson
depression and the sunspot would be disconnected and disrupted
immediately below that level, making it difficult to imagine that it
could survive as a coherent entity thereafter.

This interpretation is further supported by runs with varied efficiency
of convection, which affects the downward progression speed of the
region of convective cooling.  For the case shown in
Figs.~\ref{fig:evolution_t}--\ref{fig:evolution_fci}, we have
$\alpha=l/H_p=0.3$, which gives a heat flux density ratio of
22\%. Larger (smaller) values of $\alpha$ lead to a faster (slower)
downward extension of the convectively cooled region and to deeper
(shallower) explosion: for $\alpha=0.4$ we find an explosion depth of
about 6~Mm and a flux of 30\% of the undisturbed solar flux density,
while a tube model with $\alpha=0.2$ explodes at a depth of 2.3~Mm and
shows only 14\% of the undisturbed solar energy flux density. In both
cases, the corresponding explosion depths and times are consistent with
the downward propagation speed of the cooling flank.

\section{Discussion}
\label{Sec:discussion}

The results from our illustrative model indicate that dynamical
disconnection of emerged magnetic flux is possible at a depth of less
than about 10~Mm and within a few days after emergence if upflow
velocities as shown by simulations of thin flux tubes are considered (of
the order of a km/s at 10~Mm depth).  However, we have to keep in mind
that the assumption of a quasi-static state in the model forces us to
use a too strong initial magnetic field in the deeper layers, so that
the time evolution as shown here may not correctly represent what is
actually happening below a sunspot.  A more realistic time evolution of
the disconnection process requires a full MHD simulation.  On the other
hand, the disconnected state represents a fully self-consistent solution
with realistic surface properties and a hydrostatic stratification as
indicated by observations \citep[e.g.,][]{Beckers:1977}. Our results
show that a disconnected magneto-static sunspot model can be
consistently constructed.

We have taken the equipartition field strength with respect to the
convective velocities in the mixing-length model of the (non-magnetic)
external medium as the limit for the onset of disconnection. This is only
a rough indicator and the detailed disconnection process certainly is 
more complex. For instance, we have not considered the possible effect
of fragmentation by the interchange (fluting) instability at the
periphery of a flux tube bounded by a current sheet
\citep{Parker:1975b}. \citet{Meyer:etal:1977} have shown that buoyancy
may stabilize a sunspot only down to a few Mm depth below the solar
surface. Deeper down the destabilizing effect of the curvature of the
surface bounding the flux tube dominates. In our case, the weakening of
the field by the upflow leads to a hourglass-like shape of the sunspot,
so that buoyancy effects in the lower part actually promote the
instability. This could lead to fragmentation even before
the equipartition limit has been reached. 

The requirement for disconnection of emerged flux does not only apply to
sunspots but covers all magnetic flux in bipolar regions. We suggest
that the smaller flux tubes which form the plage and network regions are
dynamically disconnected by the same process that affects the
sunspots. In fact, using the thin flux tube approximation we find a
shallow disconnection depth (of about 2 Mm) for such structures. This
could be even smaller since a large part of the radiative energy loss of
a small tube is provided by lateral radiative heating in the surface
layers, so that the cooling of the subsurface layers proceeds more
slowly than in sunspots, leading to a more shallow disconnection. It is
well conceivable that the weakening of the field by the upflow in small
flux fragments extends up to the surface, so that all initial
connectivity is lost already during emergence or shortly thereafter. The
formation of intense fields and the patterns of flux distribution in
plages then results from the interaction with the local velocity fields
at the solar surface \citep[e.g.,][]{Voegler:etal:2005}.

In which way does the fragmented flux below the disconnected surface
flux evolve further?  Since the flux fragments are passive with respect
to convection, they will be deformed and stretched by the convective
motions, probably keeping the field strength at about the equipartion
value. However, this must not lead to a dynamical `re-connection' of the
emerged part with its magnetic roots. We conjecture that this could be
prevented by progressing fragmentation into ever thinner filaments,
owing to various instabilities and dynamical processes
\citep[e.g.,][]{Schuessler:1984,Vishniac:1995b}.  The fragmentation
proceeds until the resistive length scale is reached, for which flux
freezing is no longer valid and efficient exchange of mass and heat (by
radiation) is provided \citep{Schuessler:1987}. The magnetic flux then
has become truly passive and is kinematically transported by the motions
of the plasma, providing no further possibility for restoring the
dynamical link between the emerged flux and its the deep-lying parent
flux tube. Such a transition to passive, weak field after the explosion
of a flux tube has also been found in the numerical simulations of
\citet{Rempel:Schuessler:2001}. Sustained outflow of buoyant plasma from
the `stumps' of the disconnected parent flux tube and the horizontal
motion driven by the curvature force until a narrow U-loop remains could
also provide a mechanism for the removal of the parent flux tube from
its storage region at the bottom of the convection zone. These processes
can actually be observed in thin-tube simulations (Caligari, Rempel \&
Sch{\"u}ssler, in preparation).

\section{Conclusion}
We have shown that, in the framework of a simple magnetostatic model,
the dynamical disconnection of emerged magnetic flux from its magnetic
roots due an upflow along the field is a rather robust feature, which
occurs over a wide range of inflow velocities and within a reasonable
interval of time after emergence. The disconnection depth lies between 2
and 6 Mm for disconnection  times less than 3 days, which is consistent with
the observation that, shortly after the flux emergence phase, active
regions do no longer reflect the dynamics of their magnetic `roots' deep
in the convection zone. The disconnection depth is mainly determined by
the extension of the radiatively and convectively cooled subsurface
layer at the time of disconnection.  Radiative cooling and downflow lead
to a strengthening of the field in the upper layers and thus prevent a
loss of equilibrium at the surface, which would lead to a complete
destruction of the visible sunspot. Our results are consistent with
findings from local helioseismology with regard to the subsurface
structure of sunspots and active regions \citep[e.g.,][]{Zhao:etal:2001,
Couvidat:etal:2004}.

\acknowledgement{This work has immensely benefited from numerous
discussions and much exchange of ideas over the years with Fernando
Moreno Insertis.  Helpful comments by Yuhong Fan on a draft version of
this paper are gratefully acknowledged.  The National Center for
Atmospheric Research (NCAR) is sponsored by the National Science
Foundation.}

\bibliographystyle{aa}
\bibliography{AA-2005-2962.bbl}

\end{document}